\documentclass[a4paper,prl,twocolumn,superscriptaddress]{revtex4-1}
\pdfoutput=1

\usepackage[colorlinks,linkcolor={blue},citecolor={blue},urlcolor={blue}]{hyperref}

\usepackage{graphicx}

\usepackage{bm,amsmath}

\usepackage{latexsym}

\usepackage{verbatim}

\usepackage{color}

\newcommand{\nhat}{\hat{\bf n}}

\newcommand{\zhat}{\hat{\bf z}}

\newcommand{\nb}{\mathbf{n}}

\newcommand{\rb}{{\bf r}}

\newcommand{\qb}{{\bf q}}

\newcommand{\gb}{{\bf g}}

\newcommand{\vb}{{\bf v}}

\newcommand{\Jb}{{\bf J}}

\newcommand{\sigtens}{\mbox{\boldmath $\sigma$\unboldmath}}

\newcommand{\etatens}{\mbox{\boldmath $\eta$\unboldmath}}

\newcommand{\nablab}{\mbox{\boldmath $\nabla$\unboldmath}} 

\newcommand{\Gammab}{\mbox{\boldmath $\Gamma$\unboldmath}} 

\newcommand{\Btilde}{{\tilde B}}

\newcommand{\Ktilde}{{\tilde K}}

\newcommand{\Ctilde}{{\tilde C}}

\newcommand{\bsf}[1]{\textsf{\textbf{#1}}}

\newcommand{\beq}{\begin{equation}}

\newcommand{\eeq}{\end{equation}}

\newcommand{\bea}{\begin{eqnarray}}

\newcommand{\eea}{\end{eqnarray}}

\begin{document}

\title{Live Soap: Order, Fluctuations and Instabilities in Active
Smectics}


\author{Tapan Chandra Adhyapak}

\email{tapan@physics.iisc.ernet.in}

\affiliation{Centre for Condensed Matter Theory, Department of Physics, Indian Institute of Science, 
Bangalore 560 012 India}

\author{Sriram Ramaswamy}%

\email{sriram@physics.iisc.ernet.in}

\affiliation{Centre for Condensed Matter Theory, Department of Physics, Indian Institute of Science, Bangalore 560 012 India}

\author{John Toner}%

\email{jjt@uoregon.edu}

\affiliation{ Department of Physics and Institute of Theoretical Science, University of Oregon, Eugene, OR 97403, USA}

\date{\today}

\begin{abstract}

We construct a hydrodynamic theory of noisy, apolar active smectics, in
bulk suspension or on a substrate. Our predictions include: quasi-long-ranged
smectic order in dimension $d=2$, and long-ranged in $d=3$, extending 
previously published results to all dynamical regimes;
Kosterlitz-Thouless melting to an active nematic at high \textit{and} low
concentrations in $d=2$; nonzero second-sound speed parallel to the layers; the
suppression of giant number fluctuations by smectic elasticity; instability to
spontaneous undulation and flow in bulk contractile smectics; a layer spacing 
instability, possibly oscillatory, for large enough extensile active stresses.
 
\end{abstract}


\maketitle

Active particles \cite{schweitzer2003} in a state of orientational order
\cite{tonertusrAnnPhy2005,srannurev2010} exhibit fluctuations
\cite{tonertuPRL1995,sradititonerEPL2003,vjSCIENCE2007, chatenematicPRL2006}
and flow properties
\cite{aditisrPRL2002,VoituriezEPL2005,srmadanNJP2007,hatwalnePRL2004,tanniecristinaPRL2003,fieldingPRE2011}
that differ strikingly from those in equilibrium systems with the same spatial
symmetry.  Translationally ordered active matter has received less attention
\cite{srmadanNJP2007,sraditiSSCOM2006}.  This paper studies active systems with
spontaneously broken translation-invariance in one direction -- active smectics
-- in a broad range of dynamical regimes.  


We consider apolar systems of particles carrying an axis of orientation disposed
on average along the normal to the smectic layers (i.e., Smectics A
\cite{deGennes}), with active stresses \cite{hatwalnePRL2004} pulling in or
pushing out, i.e., contractile or extensile, along that axis. A variety of such
models with this symmetry are possible, depending on conservation laws and
geometry; we study five of these. The simplest is systems with {\it no}
conserved quantities. Such a model describes the dynamics of Rayleigh-B\'enard
roll patterns \cite{velardes, *chandrasekhar1961, *koschmieder,
*ahlers, Gollub_Langer_RMP1999}, and spontaneously layered flocks
of self-propelled
apolar entities, reproducing or dying while in motion \cite{tonerPRL2012}, 
on a substrate which serves as a momentum sink. The second is
layered flocks moving on a substrate {\it with} number conservation, but without
momentum conservation. The third is \textit{bulk} layered systems in a
background fluid with momentum conservation treated in the ``Stokesian", i.e.,
viscosity-dominated, limit appropriate for colloidal or microbial active
systems,  and the fourth is such systems \textit{confined} between no-slip
walls, where the surfaces are a momentum sink and the hydrodynamic interaction
is screened at long wavelengths.  We conclude by analyzing bulk systems in a
background fluid at wavelengths beyond the Stokesian regime, where inertia
dominates over viscosity.

Our results: (i) In all the five cases we study, smectic
order, when dynamically stable, is long-ranged in the presence of noise in
dimension $d=3$ and quasi-long-ranged in $d=2$. This reinforces and extends the
findings of \cite{sraditiSSCOM2006}. (ii) The active smectic undergoes a
transition to an active nematic as the concentration of active particles is
varied, in all five cases. In two dimensions, ``reentrance" \cite{reentrance_markoPRA1989}
{\it necessarily} occurs: the active nematic occurs at both large and small
concentration, with the active smectic at intermediate concentrations. Both
transitions, in the ``no-conservation" case, are of Kosterlitz-Thouless type
\cite{kosterlitzJPHYSC1973, *kosterlitzJPHYSC1974, *proginlowtemp}; the nature
of the transitions for the other four models is unknown.  (iii) Bulk smectic
liquid crystals in the Stokesian limit are hydrodynamically \textit{stable} to
the presence of \textit{extensile} active stresses with magnitude below a
threshold value
\cite{instability_orientable}
(iv) Active smectics, unlike their orientationally ordered counterparts
\cite{tonertuPRL1995,vjSCIENCE2007,tonertusrAnnPhy2005,srannurev2010}, have
finite concentration fluctuations, whose magnitude, however, diverges,
as activity approaches the threshold. (v) Bulk active smectics with contractile active
stresses are generically unstable without threshold to spontaneous undulations
\cite{HelHar_active}
and flow. (vi) Confined active
smectic suspensions are in general stable at long wavelengths for small enough
active stresses of either sign.  Beyond a threshold value of activity the
confined system too undergoes an instability, which is likely to be oscillatory
for the extensile case.  {(vii)} For stable bulk active smectics, the speed of
the smectic second sound mode is nonzero for propagation parallel to the
smectic layers. Our findings apply to active smectics in a wide range of
settings including vibrated granular layers \cite{vj2006} and the
Rayleigh-B\'enard problem \cite{velardes, *chandrasekhar1961, *koschmieder,
*ahlers}. Agitated 2DEGs \cite{zudovPRB2001,*maniNATURE2002,*aliceaPRB2005,*foglerPRL2000,*radzihovskyPRL2002}, 
where Coulomb and magnetic-field
effects enter, will be discussed elsewhere\cite{tapan_tobepublished}.

We begin with the simplest case: active elements whose number is {\it not}
conserved, spontaneously condensed into a uni-directional periodic
structure, i.e., a smectic A, with mean layer normal $\nhat$ along
$\zhat$. The only hydrodynamic field in this case is the broken symmetry
variable $u$ giving displacements of the layers along $\zhat$. This model  also
describes Rayleigh-B\'enard stripes \cite{velardes,
*chandrasekhar1961,
*koschmieder, *ahlers}, where the modulated field
is the local temperature, which is not a conserved quantity.
This case was dealt with briefly in \cite{sraditiSSCOM2006}. 
The hydrodynamic, long-wavelength model for
the dynamics of the $u$-field, retaining terms permitted by symmetry, including
$u \to -u, \, z \to -z$ \cite{notepolar}, to leading order in gradients and in
powers of $u$, reads 
\begin{eqnarray}
\partial_t u = \Btilde \partial_z^2 u + D \nabla_{\perp}^2 u -
\Ktilde\nabla_{\perp}^4 u + f^u, 
\label{u no cons} 
\end{eqnarray} 
where $f^u$ is a Gaussian, zero-mean spatiotemporally white noise with variance
$2\Delta$.
The term with coefficient $D$ \cite{active_tension}
is forbidden by rotation-invariance of the free energy in
an \textit{equilibrium} smectic without an aligning field. It is however
permitted here simply because rotation-invariance \textit{at the level of the
equation of motion}, which is all one can demand in an active system, does not
rule it out. Its physical content is that layer curvature produces a local
vectorial asymmetry which must cause directed motion of the layers as this is
a driven system. Symmetry does not fix the sign of
$D$. An undulation instability \cite{helfrich1970,*hurault1973} arises if
$D<0$. For positive $D$ the linearity and spatial homogeneity of (\ref{u no cons}) makes  it straightforward to show via spatial Fourier transform
that the variance $\langle |u(\qb, t)|^2\rangle =
\Delta/(\tilde{B}q_z^2 + D q_{\perp}^2)$ in Fourier space for small wavevectors
$\mathbf{q} =
(\mathbf{q}_{\perp},q_z)$. This implies \cite{sraditiSSCOM2006} a 
real-space variance $\langle u^2\rangle =
\int_{\mathbf{q}}\langle |u(\qb, t)|^2\rangle$ that is finite in $d=3$,
corresponding to long-range smectic
order; in $d=2$, $\langle u^2 \rangle \sim \log L$ for system size $L$,
corresponding to quasi-long-range order. This establishes result (i) for the
simplest case.

Ignoring the $\Ktilde$ term, which is irrelevant at long distances, rescaling
$\rb_\perp=\rb_\perp'(D/\Btilde)^{1/2}$, 
$z' = z$, and expressing $u$ in terms of
the angle $\theta(\rb')\equiv (2\pi/a)u(\rb)$ where $a$ is the layer spacing, we
can rewrite (\ref{u no cons}) in the form
\beq
\label{thetaeq}
\partial_t \theta = \Btilde \nabla'^2\theta + f^\theta,
\eeq 
with rescaled noise statistics 
\beq
\label{thetanoise}
\langle f^\theta(\mathbf{r}',t) f^{\theta}(\mathbf{0},0)\rangle =  \left({2 \pi
\over a}\right)^2 \Delta \left({\Btilde \over D}\right)^{d-1 \over 2}
\delta(\mathbf{r}'_{\perp})
\delta(z')\delta(t).
\eeq 
Defining 
$\kappa \equiv  {\Btilde^{(3-d)/ 2} D^{(d-1)/2}a^2/(2
\pi^2\Delta)}$
it can be shown that the steady-state probability distribution
for $\theta$ implied by (\ref{thetaeq}), (\ref{thetanoise}), is
$\exp [(-\kappa/2) \int d^dr' (\nabla' \theta)^2]$, identical to that for a
thermal equilibrium $XY$ model with a stiffness/temperature ratio $\kappa$.

\begin{figure}[h]
\includegraphics[height=3.8cm]{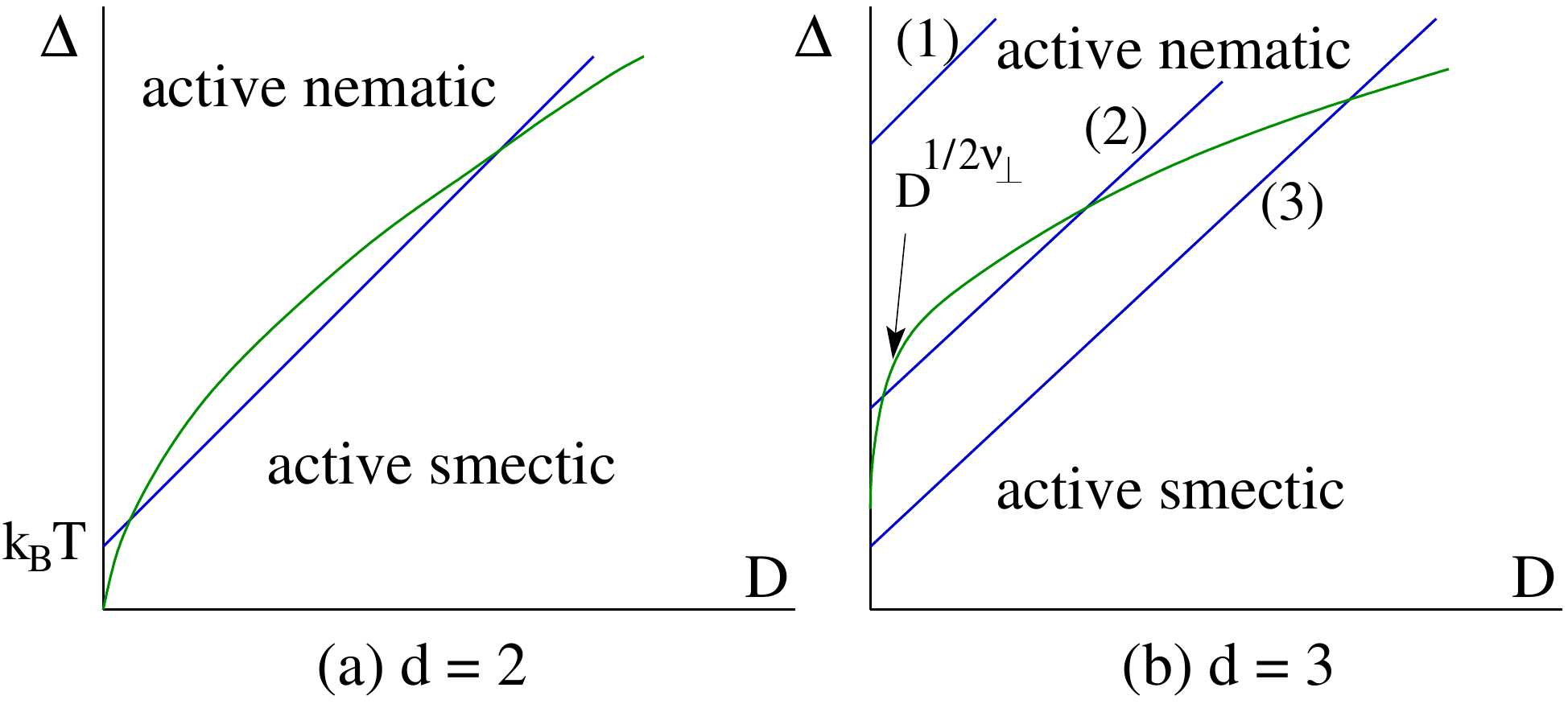}
\caption{Phase diagram of active smectics, in (a) $d = 2$ 
and (b) $d = 3$.} 
\label{phasediag}
\end{figure}

This equivalence to an equilibrium XY model implies
that topological defects (i.e., dislocations) in an active smectic in dimension $d=2$ unbind, driving the system into the active nematic phase,
\cite{kosterlitzJPHYSC1973, *kosterlitzJPHYSC1974, *proginlowtemp} when $\kappa
= 2/\pi$, i.e., when 
$2\pi^2\Delta /a^2 (\Btilde D)^{1/2}=\pi / 2~.  $
This locus is plotted in the $\Delta$-$D$ plane in figure \ref{phasediag}(a). 


Since it is a purely active effect, we expect $D\propto c_0$. We also expect the noise strength $\Delta$ to  get an active
contribution proportional to $c_0$, and  a $c_0$-independent thermal
contribution  proportional to $k_BT$. 
Hence, varying $c_0$ maps out  a straight line with positive intercept on the
$\Delta$-axis in the $\Delta$-$D$ plane, as illustrated in figure \ref{phasediag}(a).  As is clear from that figure,
this experimental locus can  only enter the active smectic region
by crossing the active smectic to active nematic phase boundary twice. Hence our conclusion that re-entrance is inevitable in
two dimensions for these systems. 

In three dimensions, the situation is quite different, because {\it
equilibrium} smectics are {\it stable} against dislocations in $d=3$, since
dislocations are line defects that remain bound even in the absence of the $D$
term in (\ref{u no cons}) \cite{jt1stpaperPRB1981}
As a result, the active
smectic- active nematic phase boundary does {\it not} go all the way down to
$\Delta=0$ at $D=0$ in $d=3$. It {\it does}, however,  develop an infinite
downward slope at $D=0$, as can be deduced by 
the following argument: for small
$D$, as the transition is approached, the system will act like an equilibrium
($D=0$) system until the $D$ term in (\ref{u no cons}) becomes comparable to
the $\Ktilde$ term at in-plane wave vectors $q_\perp\sim\xi^{-1}_\perp$, where
$\xi_\perp$ is the equilibrium, in-plane correlation length for smectic order.
This condition leads to 
$D/\xi^2_\perp \sim \Ktilde/\xi^4_\perp$,
which implies 
\begin{eqnarray} 
\xi_\perp\sim\sqrt{{\Ktilde / D}}.
\label{xi3d}
\end{eqnarray} 

Near the equilibrium $AN$ transition, 
$ \xi_\perp\propto|T-T_{AN}|^{-\nu_\perp}$ where $\nu_\perp$ is the
``thermodynamic'' equilibrium correlation length exponent in the $\perp$ direction; 
see reference \cite{jtPRB1982} for a further discussion.
Since  $\Delta$ plays the role of temperature here, (\ref{xi3d}) leads
to a shift 
 $\propto D^{1 / 2\nu_\perp}$ in the critical $\Delta_{cr}$.
Both theory \cite{jtPRB1982}
and experiment \cite{garlandPRE1993} 
find $1/2\nu_\perp<1$, 
so the 
phase boundary in figure \ref{phasediag}(b)  has infinite slope as $D\rightarrow 0$. 

The 
locus in the $\Delta$-$D$ plane mapped out by varying $c_0$ remains a 
straight line;
now, however, re-entrance, though still obviously a possibility (e.g., for locus
$2$ in figure \ref{phasediag}(b)),
can  be avoided, as on locus $3$ in figure \ref{phasediag}(b).

As in $d=2$, in $d=3$ the active smectic to active nematic transition is in the
XY model universality class for the model with no conservation laws.  While the
other four dynamical models we study in this paper are {\it not} equivalent to
{\it any} equilibrium XY model,  they  have exactly the same scalings of their
equal time $u$-$u$ correlations as this simplest model.  Hence, we expect
similar phase diagrams, with the 
critical $\Delta_{cr}\propto D^{1/2}$ 
for small
$D$ in $d=2$. However, the universality class of the phase transition  in these
other models may be different.

We now turn to our second  model, in which the number of active particles {\it
is} conserved, but momentum is not. Now the concentration $c$ of active
particles with mean $c_0$ and $\delta c \equiv c-c_0$, joins the broken symmetry variable $u$ as a hydrodynamic field.
For {\it apolar} phases, the equations of motion must be unchanged if 
$u\rightarrow-u$ and $z\rightarrow -z$ simultaneously. The $c$-dependent term in the $u$ equation of motion that is lowest order in spatial derivatives that respects this symmetry is  $\propto\partial_z c$; hence, the equation of motion for $u$ is:
\begin{eqnarray}
\label{ueqn2}
\partial_t u =  \Btilde \partial_z^2 u + D \nabla_{\perp}^2 u -
\Ktilde\nabla_{\perp}^4 u + \Ctilde \partial_z \delta c +f^u~, \label{eqn:u c
cons}
\end{eqnarray}
with $f^u$ having the same statistics as those in (\ref{u no cons}).

The equation of motion for the concentration $c$ must, due to  conservation of total number of particles, be expressible in the form: 
$\partial_t c=-\nablab\cdot \Jb^c$. Gradient expanding the current subject to the symmetry constraints  gives:  
\begin{eqnarray}
\label{eqn:current}
\mathbf{J}_c =&&-\hat{\bf z} [(A_z -W^c)\partial_z \delta c +
W^cc_0\nabla^2_{\perp} u +
C_{zz} \partial_z^2 u ]  \nonumber \\
        &-&  \nablab_{\perp}[A_{\perp}\delta c +
(C_{\perp z}+W^cc_0)
\partial_z u] 
+\mbox{\boldmath $f$\unboldmath}^c
\end{eqnarray}
where the Gaussian noise  $\mbox{\boldmath $f$\unboldmath}^c =
(\mbox{\boldmath $f$\unboldmath}^c_{\perp}, f^c_z)$ has statistics
\beq
\langle f^c_i(\rb,t) f^c_j(\mathbf{0},0)\rangle=
(\Delta^c_\perp\delta^\perp_{ij}+\Delta^c_z\delta^z_{ij}
)\delta^d(\rb)\delta(t)  .
\label{c noise correlation}
\eeq
In (\ref{eqn:current}) we have included an active current
\cite{sradititonerEPL2003} 
$W^c\nablab\cdot(c\nb\nb)$ with
$\hat{\nb}=(\zhat-\nablab u) /|\zhat-\nablab u|$, where $W^c$ is
a phenomenological coefficient.
In an {\it equilibrium} two component smectic, (\ref{eqn:current}) and
(\ref{c noise correlation}) 
would hold, but with the constraints $W^c=0$ 
and $C_{\perp z} /
C_{zz} = A_{\perp}/A_z=\Delta^c_{\perp}/\Delta^c_z$.

The two key results that emerge from (\ref{ueqn2})-(\ref{c noise
correlation}) are that, for $q \to 0$ and for all directions of $\mathbf{q}$,
the equal-time correlators
$\langle |u_{\bf q}|^2 \rangle \propto {1/q^2}$
for {\it all} directions of $\bf{q}$, and that $\langle |\delta c_{\bf q}|^2
\rangle$ is finite.
 As before, this $q^{-2}$ scaling of $u$ fluctuations 
implies 
translational order is quasi-long-ranged in $d=2$, and long-ranged  in $d=3$.
The finite concentration fluctuations, which result from the smectic elasticity
$\Btilde$, imply the absence of giant number fluctuations  (our result (iv)).

We next consider active smectics suspended in a fluid medium. The total momentum
of suspended particles and ambient fluid is conserved; the corresponding
momentum density $\gb$ is therefore slow and hydrodynamic. The other
hydrodynamic fields  $u$  and  $c$ remain as well, of course. We'll assume
overall  incompressibility, so that total (particle + fluid) mass density
$\rho=\rho_0=$ constant and $\nablab\cdot \vb=0$, where $\vb \equiv \gb / \rho$
is the velocity field. 

Conservation of total momentum reads $\partial_t\gb=-\nablab\cdot\sigtens$, with
a linearized stress tensor
\beq
\label{stresstot}
\sigtens = p\bsf{I} -\etatens (\nablab \vb+ \nablab \vb^T) 
+\sigtens^{(el)}
+\sigtens^a + \sigtens^N~,
\eeq
with $p$  the fluid pressure, $\etatens$  the viscosity tensor, and the  elastic
 force density $-\nablab \cdot \sigtens^{(el)}= -\nb \delta F / \delta u$, with 
\beq
\label{elasticity}
F = {1 \over 2}\int d^d x \left[B(\partial_z u)^2 +K(\nabla_\perp^2 u)^2 + A(\delta c)^2 + 2C\delta c\partial_z u\right].
\eeq
Here $B$  and $K$ are layer compression and bend moduli respectively,   $A$ the osmotic modulus, and $C$  a cross-coupling. The active stress \cite{aditisrPRL2002, srannurev2010} 
$\sigtens^a = -W c \nb \nb$, with negative and positive $W$ corresponding respectively to extensile and contractile stresses, while $\sigtens^N$ is noise. That $\sigtens^a\propto c$ follows because
$W$ is   the activity {\it per particle}.
As in any smectic A, 
the layer normal $\nb$ is geometrically locked 
to the displacement field: 
$\delta \nb \equiv \nb - \zhat \simeq -\nablab_{\perp}u$.

The resulting equation of motion for $ \mathbf{v}$, linearized in $\vb$, $u$ and $\delta c = c - c_0$, with $c_0$  the mean concentration, reads 
\begin{eqnarray}
	\rho_0 \partial_t \mathbf{v} &=& -\nablab p +\zhat[B \partial_z^2 u - K\nabla_{\perp}^4 u + (C+W)\partial_z \delta c]\nonumber \\  && -Wc_0(\zhat \nabla_{\perp}^2 u+\partial_z\nablab_{\perp}u) - \Gammab \cdot \mathbf{v}  +\mathbf{f}^v, \label{eqn:rhov}  
\end{eqnarray}
\noindent  
where 
$-\Gammab \cdot \mathbf{v} = \nablab\cdot(\etatens \nablab \mathbf{v})$,
$\mathbf{f}^v = \nablab \cdot \sigtens^N$ is a momentum-conserving noise, and
$\langle \sigma_{ij}^N(\mathbf{0},0) \sigma_{kl}^N(\mathbf{r},t) \rangle = 2
\Delta_{ijkl} \delta(\mathbf{r})\delta(t)$. $\Delta_{ijkl}$ is uniaxial,
symmetric in $ij$ and $kl$ \textit{and under interchange of} $ij$ with $kl$, and
thus has five independent components in  $d=3$. In thermal equilibrium
$\Delta_{ijkl} \propto \eta_{ijkl}$, but not in general nonequilibrium
systems \cite{grinstein_lee_sachdev}.
We take  $\Delta_{ijkl}=\Delta_v
(\delta_{ik}\delta_{jl}+\delta_{il}\delta_{jk})$ and
$\eta_{ijkl}=\eta\delta_{jk}\delta_{il}$. 
The linearized hydrodynamic equation of motion for  $u$ is  
\begin{eqnarray}
\partial_t u = v_z+ \Btilde \partial_z^2 u + D \nabla_{\perp}^2 u -
\Ktilde\nabla_{\perp}^4 u + \Ctilde \partial_z \delta c +f^u; 
\nonumber \\
\label{eqn:u}
\end{eqnarray}
where the noise $f^u$ has statistics as in 
(\ref{u no cons}) \cite{nocrosskin}.

The equation of motion for 
$c$ 
is: $\partial_t c=-\nablab\cdot \mathbf{J}^c$ with 
$\mathbf{J}_c$
given
by equation (\ref{eqn:current}).

For simplicity, we take
$\Delta^c_{\perp} = \Delta^c_z
\equiv \Delta^c$,
$C_{\perp z} = C
\Delta^c = C_{zz}$,  and $A_{\perp} = A \Delta^c = A_z $, where $A$ and $C$ are as
in (\ref{elasticity}),
and define $C \Delta^c \equiv E, \, A \Delta^c \equiv G$. 
Activity now enters only through the $W$ terms in (\ref{eqn:rhov}).

To study the Stokesian limit, we neglect inertia and acceleration,  impose incompressibility $\nabla \cdot \mathbf{v} = 0$ in (\ref{eqn:rhov}), and solve for $\mathbf{v}$ in terms of $u$ and $c$. Inserting the result in (\ref{eqn:u}), and defining $\Phi \equiv -\partial_z u$, we find that the spatial Fourier transforms $\delta c_{\bf q}, \, \Phi_{\bf q}$ obey 
\bea 
\label{eqn:u:eff}
\partial_t{\Phi_{\bf q}}=  -M_{\mathbf{q}}\{[B q_z^2 + W c_0 (q_z^2
-q_{\perp}^2) +K q_{\perp}^4 ] \Phi_{\bf q}  \nonumber&&\\
-(C+W) q_z^2\delta c_{\bf q}\}  + [\partial_t \Phi_{\bf q}]_P
- i q_z(f^u_{\mathbf{q}} + M_{\bf q} f^v_{z \mathbf{q}})&& 
\eea
\beq
\partial_t\delta c_{\bf q} = (E q^2 + 2W^cc_0q_\perp^2) \Phi_{\bf q} - (G
q^2-W^cq_z^2)\delta c_{\bf q} -
i \mathbf{q} \cdot \mbox{\boldmath $f$\unboldmath}_{\mathbf{q}}^c
\label{eqn:c:eff}
\eeq
where $M_{\mathbf{q}} \equiv q_{\perp}^2 / \eta q^4$, and $ [\partial_t
\Phi_{\bf q}]_P$ summarizes the ``permeative'' $\Btilde, \Ktilde, \Ctilde, D$
terms from (\ref{eqn:u}), which are of higher order in wavenumber than those
shown explicitly in (\ref{eqn:u:eff}).


Suppose $B > CE/D = C^2 /A$, so that when activity $W = 0$ the
smectic state is stable. Let $|W | > C > 0$; a similar analysis holds for $C
< 0$. At small $q$, where $[\partial_t \Phi_{\bf q}]_P$ is
negligible, it is clear from (\ref{eqn:u:eff}) that negative (i.e., extensile)
$W$, can lead to an instability with $\bf{q}$ along $z$, i.e., 
a modulation in layer spacing.  However, the layer compression
modulus $B$ always stabilizes this when $(B-|W|c_0) >0$. Thus,
the system is stable for small enough $|W|$, establishing our result (iii).

For contractile active stresses $W>0$, we see from (\ref{eqn:u:eff}) and
(\ref{eqn:c:eff}) that the most unstable modes have $\bf{q}$ in the $\perp$
direction, in which  neither the layer compression elasticity nor the coupling
to the concentration act. Hence,  the instability threshold for $W\to 0$  in the
limit of large system size, as in \cite{aditisrPRL2002,VoituriezEPL2005}. The
instability causes splay and self-generated flow, as in active nematics
\cite{aditisrPRL2002,VoituriezEPL2005}. For smectics, this is a spontaneous
version of the Helfrich-Hurault \cite{helfrich1970,*hurault1973,deGennes}
undulation instability.

We turn next to the effects of confinement.
Consider an active 
smectic  with layers normal to the $z$ direction, confined between no-slip walls parallel to the $xz$ plane, 
a distance $\ell$ apart. We start with (\ref{eqn:rhov}) but with both
$\mathbf{f}^v$ and $\Gammab$ $\sim \eta /\ell^2 \equiv \Gamma$ nonzero at zero
wavenumber because the walls are a momentum sink. Solving the modified
(\ref{eqn:rhov}) for the (now) fast variable $\mathbf{v}$ in terms of the slow
$u$ and $c$ and inserting the result in (\ref{eqn:u}) yields
\begin{eqnarray}
\label{eqn:u:eff:confine}
\partial_t{\Phi_{\bf q}}= &-&(\bar{B}q_z^2 + \bar{D}q_x^2 +
\bar{K}q_x^4)\Phi_{\bf q} + \bar{C}q_z^2 \delta c_{\bf q} \nonumber \\
&-& i q_z(f^u_{\mathbf{q}} + M_{\mathbf q}f^v_{z \mathbf{q}}) ,
\end{eqnarray}
while equation (\ref{eqn:c:eff}) continues to hold.
Here $\bar{B}=\tilde{B} + (B+Wc_0) M_\mathbf{q}$, $\bar{D}=D-Wc_0
M_\mathbf{q}$, $\bar{K}= \tilde{K}+KM_\mathbf{q}$, and
$\bar{C}=\tilde{C}+(C+W) M_\mathbf{q}$.
The wave vector $\mathbf{q}$ now lies in the $xz$
plane and the mobility $M_\mathbf{q}=(q_x^2/q^2)(1/\Gamma)$, in contrast to the
bulk system, does not diverge at small $q$, so that terms involving $M_{\bf q}$,
which arise from the (screened) hydrodynamic interaction, are of the same order
in wavenumber as the permeative terms $[\partial_t \Phi_{\bf q}]_P$. 
Ignoring the concentration field, we see that the relaxation rate of layer
displacements with wavevector in the $x$ direction is now proportional to $[D -
(c_0W/\Gamma) q_x^2/q^2] q_x^2$ for $q_x \gg q_z$ and $\{\tilde{B} + [(B + c_0
W) / \Gamma] q_x^2/q^2\}q_z^2$ for $q_z \gg q_x$. Thus, there is a range of
parameters for which the active smectic is stable (result (vi)). For other
parameter ranges instabilities occur; e.g.,  
if $D=0$ in (\ref{eqn:u}), an undulation instability occurs 
for 
$\bf{q}\perp z$,  despite confinement. 

The instability that arises 
 in the extensile ($W<0$)
case when $|W|>B$ is interesting. 
Equations
(\ref{eqn:u:eff:confine}) and (\ref{eqn:c:eff}) have the same form as the
linear part of the Fitzhugh-Nagumo \cite{fitz1961, *nagumo1962, murray} equation, which exhibits sustained oscillations under
rather general conditions. 
We speculate that 
such oscillations could also occur here; i.e., a breathing smectic. We will explore this in future work.

We now turn to  fluctuations 
in the bulk Stokesian limit. 
For $q_z\to0$, $\delta c$ drops out of (\ref{eqn:u:eff}), and it is then easy to
check for extensile activity that $\langle |u_{\bf q}|^2
\rangle = 2 \Delta_v /|W|c_0 q_{\perp}^2$. Somewhat more tedious algebra, which
we'll present elsewhere, shows  that $\langle |u_{\bf q}|^2 \rangle \sim 1/q^2$
for all directions of $\mathbf{q}$. We can also see from (\ref{eqn:u:eff}) that,
in the Stokesian regime, time correlations of $u$ decay   at a nonzero rate in all directions for $q \to 0$.
These facts 
imply that, in bulk active smectic suspensions as well, the coupling to $u$ via
the $E$ term in (\ref{eqn:c:eff}) won't lead to diverging concentration
fluctuations in general. $\langle |\delta c_{\bf q}|^2 \rangle$ does diverge,
however, upon approaching the extensile instability, with a
correlation length $\sim (B-|W|c_0)^{-1/2}$.

We conclude with second sound.
Leaving
the steady Stokesian regime, taking acceleration ($\partial_t \mathbf{v}$) into
account, and working at long wavelengths where viscosity is negligible, makes the coupled
dynamics of $\mathbf{v}$ and $u$, for wavevectors in the plane of the layers 
$\rho_0 \partial_t \mathbf{v} = -\nabla p  -Wc_0 \zhat \nabla_{\perp}^2 u; \, \partial_t u = v_z; \, \nabla \cdot \mathbf{v} =0$, can readily be seen to give sound waves with a speed $\sqrt{-W c_0/\rho_0}$.

In conclusion, we have constructed the dynamical equations for active
smectics, both in bulk suspensions and in confined systems in contact with a
momentum sink. Our theory is generic, applicable to any driven system with
spontaneous stripe order. We show, extending \cite{sraditiSSCOM2006}, that
noisy active smectic order is long-ranged in dimension $d=3$ and
quasi-long-ranged in $d=2$ for all dynamical regimes, and that active
smectic suspensions have a nonzero second sound speed parallel to the layers.
For $d=2$ we predict a Kosterlitz-Thouless transition from active nematic to
active smectic, with a re-entrant nematic at low concentration. We show that
smectic elasticity suppresses the giant number fluctuations and extensile
instabilities that occur in active nematics, but that bulk contractile systems
exhibit an active undulation instability. Active extensile stresses, if
strong enough, give rise to a ``breathing" instability which is likely to be
oscillatory. Our results should apply to a wide range of active systems,
including horizontal layers of granular matter agitated vertically or fluids
heated from below. 

We look forward to detailed experimental tests of our predictions.

We are grateful to R.A. Simha for useful discussions, and the Active Matter
workshop of the Institut Henri Poincar\'e, Paris, the Lorentz Center of Leiden
University (SR and JT), the Initiative for the Theoretical Sciences at The
Graduate Center of CUNY and the MPIPKS, Dresden (JT), for support and
hospitality while this work was underway. TCA acknowledges support from the
CSIR, India, SR from the DST, India, through a J.C. Bose grant and Math-Bio
Centre grant SR/S4/MS:419/07, and JT from the U.S. National Science
Foundation through awards \# EF-1137815 and 1006171.

%


\begin{thebibliography}{47}
\bibitem [{noc()}]{nocrosskin}%
  \BibitemOpen
  \href@noop {} {}\bibinfo {note} {For simplicity, we have assumed that there
  is no cross kinetic coefficient coupling $u$ and $c$ in the absence of
  activity. The equilibrium limit in the presence of off-diagonal Onsager
  coefficients is somewhat more complicated.}\BibitemShut {Stop}%
\end{thebibliography}

\end{document}